\documentclass[prb,twocolumn,showpacs,preprintnumbers,amsmath,amssymb]{revtex4}

\usepackage{color}
\usepackage{graphicx}
\usepackage{dcolumn}
\usepackage{bm}

\preprint{submitted to Journal of Advanced Physics}

\begin{document}

\title{First-principles computed electronic and magnetic properties of zincblende alkaline-earth  pnictides}

\author{K. \"Ozdo\u{g}an}\email{kozdogan@yildiz.edu.tr}
\affiliation{Department of Physics, Yildiz Technical University,
34210 \.{I}stanbul, Turkey}

\author{I. Galanakis}\email{galanakis@upatras.gr}
\affiliation{Department of Materials Science, School of Natural
Sciences, University of Patras,  GR-26504 Patra, Greece}

\date{\today}

\begin{abstract}
Employing first-principle electronic structure calculations, we
study the magnetic and electronic properties of the XY (X= Mg, Ca,
Sr and Y= N, P, As, Sb) compounds crystallizing in the zincblende
structure. The Ca and Sr alkaline-earth metal monopnictides are
found to be half-metallic with a total spin magnetic moment per
formula unit of 1.0 $\mu_B$. In the case of the Mg alloys the p-d
hybridization effect is much weaker and only MgN is a half-metal.
Electron counting of the bands explains the Slater-Pauling
behavior exhibited by the total spin magnetic moment. We also
study for these alloys the effect of deformation taking into
account both the cases of hydrostatic pressure and
tetragonalization keeping constant either the in-plane lattice
parameters or the unit cell volume. Even large degrees of
deformation only marginally affect the electronic and magnetic
properties of these alloys. Finally, we show that this stands also
for the rocksalt structure. Our results suggest that
alkaline-earth metal monopnictides are promising materials for
magnetoelectronic
applications.
\end{abstract}


\maketitle
\section{Introduction}\label{sec1}

Half-metals are among the most studied magnetic materials
due to their potential application in spintronic
devices.\cite{Zutic,Felser,Zabel} These unusual materials present
metallic properties for one spin-direction while they behave like
semiconductors for the other spin-direction eventually leading to
100\%\ spin-polarization at the Fermi level.\cite{Katsnelson}
This peculiar property is exhibited by several transition-metal
alloys but lately half-metallic ferromagnetism has been predicted
for several alloys which do not include transition-metal atoms.
Such alloys are widely known in literature with various names like
$d^0$ magnets, $p$-ferromagnets or sp-electron
ferromagnets.\cite{Volniaska10} These compounds have the advantage
of being energy-efficient for applications since they create
weak external magnetic fields and thus lead to minimal energy
losses.  There are several ways to create sp-electron ferromagnets
and an extensive review is given in Ref. \onlinecite{Volniaska10}.

One of the most promising routes to half-metallic $sp$-electron
ferromagnets is the growth of I/II-IV/V nanostructures in
metastable lattice structures similar to the case of
transition-metal pnictides and chalcogenides in the metastable
zincblende structure.\cite{ReviewCrAs} Several studies to this
research direction have appeared following the pioneering papers
published by Geshi et al\cite{Geshi04} and Kusakabe et
al\cite{Kusakabe04} who have shown using first-principles
calculations that CaP, CaAs and CaSb alloys present half-metallic
ferromagnetism when grown in the zincblende structure. Moreover
they have shown that the gap appears due to a p-d hybridization
mechanism which is similar as we will discuss in the next
paragraph for all half-metallic ferromagnetic I/II-IV/V alloys in
all three zincblende, wurtzite and rocksalt metastable structures.
The spin magnetic moment in these alloys follows a Slater-Pauling
behavior with the total spin magnetic per formula unit in $\mu_B$
being 8 minus the number of valence electron in the unit cell:
$M_t=8-Z_t$. The eight stems from the fact that in the
majority-spin band we have exactly 4 occupied states (one s and
three p states) and the rest of the electrons occupy exclusively
the minority-spin states. The same reasoning stands even for
Heusler compounds like GeKCa and SnKCa which contain no transition
metals and have been predicted to exhibit half-metallic
ferromagnetism.\cite{Chen11} Evidence of the growth of such
nanosctructures has been provided by Liu et al who have reported
successful self-assembly growth of ultrathin CaN in the rocksalt
structure on top of Cu(001).\cite{Liu08}

\section{Overview of recent literature}\label{sec2}

Following the Refs. \onlinecite{Geshi04} and
\onlinecite{Kusakabe04} mentioned in the previous paragraph  a lot
of theoretical studies on sp-electron ferromagnets have appeared based on
first principles calculations and we will give a short overview of
them. A general trend is that the rocksalt (RS) structure is more
stable than the zincblende (ZB) or wurtzite (WZ) structures, in
all cases the Slater-Pauling (SP) rule for the total spin magnetic
moment is obeyed and the spin magnetic moment is almost entirely
concentrated to the anionic sites and the interstitial region
while the cations show a spin moment one or two orders of
magnitude smaller (note that the occupied states are almost
exclusively of anionic s- and p-character). Also in most studied
cases half-metallicity is conserved for large variations of the
lattice constant assuming hydrostatic pressure.

We will start our overview from the alkali-metal based alloys. In
2006 Sieberer and collaborators predicted half-metallicity for LiN
and Na(K)P(As) in the zincblende structure with a total spin
magnetic moment of 2 $\mu_B$ per formula unit
(f.u.)\cite{Sieberer06} and in 2008 Zhang confirmed the
half-metallicity also for the alkali-metal carbides (Li,K,Na)C
with a total spin moment of 3 $\mu_B$ in accordance to the SP
rule.\cite{Zhang08} After, Zberecki and collaborators have
demonstrated that the rocksalt is the most stable structure for
alkali-metal nitrides with respect to the zincblende and wurtzite
structures.\cite{Zberecki09} These results were independently
confirmed by Yan,\cite{Yan12} and by Gao and collaborators who
have also shown that among the alkali metal phosphides and
arsenides, only KP and KAs are half-metallic
ferromagnets.\cite{Gao09} Finally, Gao et al have also studied the
alkali-metal sulfides (Li,Na,K)S which exhibited a total spin
magnetic moment of 1 $\mu_B$ as expected by the SP rule for both
the ZB and the energetically-favored RS structures.\cite{Gao11}

Gao and collaborators have studied a series of alkaline-earth
metal compounds with the IVth column elements and found that all
CaSi, CaGe, CaC, SrC and BaC in the ZB structure are
half-metals.\cite{Gao07,Gao07c} Gao and Yao have, then, studied
SrC and BaC in more detail and have shown that the RS structure in
the most stable against the other metastable structures and found
that both RS-SrC and RS-BaC are half-metals with the Curie
temperature of SrC exceeding the room temperature (485
K).\cite{Gao07b} Dong and Zhao have shown that the alkaline-earth
metal carbides retain their half-metallicity up to a hydrostatic
pressure of 32GPa (the volume is reduced by 30\%),\cite{Dong11}
Zhang et al have studied ScC in the WZ structure,\cite{Zhang08b}
and finally Gao  have shown that the (110) surface of ZB-CaC keep
their half-metallic character contrary to the (001) surfaces.
\cite{Gao09b} Also Verma and collaborators have studied the
alkaline-earth metal silicides and found that only the
ZB-structure is half-metallic contrary to the more-stable RS
ferromagnetic structure.\cite{Verma10}

\begin{figure}
\includegraphics[width=\columnwidth]{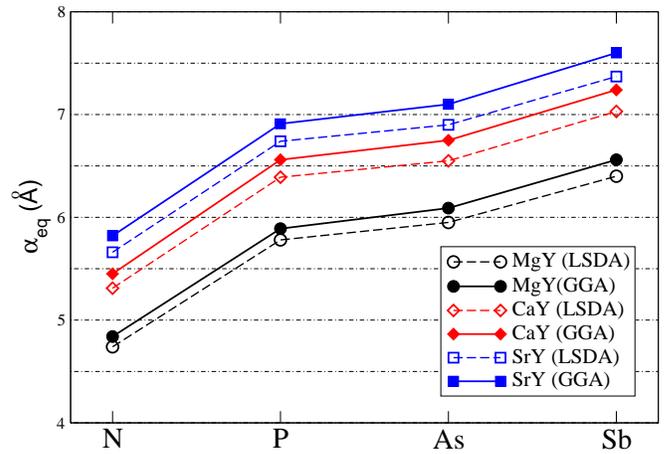}
\caption{(Color online) Calculated equilibrium lattice constant in
\AA ngstr\o m of several alkaline-earth  pnictides assuming
the zincblende structure. Total energy calculations have been
performed using both the local-spin density (LSDA) and generalized
gradient (GGA) approximations to the exchange-correlation
potential in conjunction to the full--potential nonorthogonal
local--orbital minimum--basis band structure method (FPLO).}
\label{fig1}
\end{figure}

Most of the attention has been focused on the alkaline-earth metal
(IInd column) compounds with the Vth-column elements. As we have
already mentioned half-metallicity has been predicted for the CaP,
CaAs and CaSb ZB-alloys.\cite{Geshi04,Kusakabe04} Sieberer et al
studied all possible II-V combinations in the ZB and WZ structures
and found that all alloys containing Ca, Sr and Ba are
half-metallic while only MgN was half-metallic between the
Mg-based compounds.\cite{Sieberer06} It was also shown in Ref.
\onlinecite{Sieberer06} that the ferromagnetic state is
energetically preferable to both the non-magnetic and the
antiferromagnetic configurations. Yao et al have shown that the
ZB-alloys containing Bi from the Vth column are also half
metals,\cite{Yao06} but later on Li and  Yu demonstrated that the
inclusion of spin-orbit coupling destroys half-metallicity in
these Bi-alloys.\cite{Li08} Volnianska and Boguslawski, as well as
Geshi and collaborators have studied the alkaline-earth metal
nitrides and have shown that the RS is the most stable structure
with formation energies of about -11 eV per unit
cell.\cite{Volniaska07,Geshi07} Gao et al have shown that among
the RS alloys containing Ca, Sr or Ba as a cation and N, P or As
as an anion only the nitrides are stable half-metallic
ferromagnets with a total spin magnetic moment of 1 $\mu_B$ and
cohesive energies about -9 eV per formula unit.\cite{Gao08}
Droghetti and collaborators have shown that RS-MgN is in verge of
the half-metallicity and suggested that MgN inclusion upon the
N-doping of MgO should lead to a material suitable for magnetic
tunnel junctions.\cite{Droghetti09} The most recent studies on
nitrides concern the Curie temperature which was found to be well
above the room temperature for CaN and SrN in both the ZB and RS
structures,\cite{Laref11,JAP}
 and the RS-CaN/ZB-InN and RS-SrN/ZB-GaP
(111) interfaces which were found to retain half-metallicity only
when the interface is made up from Ca-N or N-In atoms in the first
case and N-Ga in the second case.\cite{Gao11b}

\begin{table*}
\caption{Equilibrium lattice constants in \AA ngstr\o m determined
through total energy calculations employing the FPLO method within
both the LSDA and GGA approximations for all zincblende XY
compounds under study. With $m^\mathrm{X}$ and $m^\mathrm{Y}$ we
present the atom-resolved spin magnetic moment in $\mu_\mathrm{B}$
at both the cation (X) and anion (Y) sites at the corresponding
equilibrium lattice constants, as well as the total spin magnetic
moment ($m^\mathrm{total}$) per formula unit. The Mg-based alloys,
with the exception of MgN, have converged to non-magnetic ground
states.} \label{table1}
\begin{ruledtabular}
 \begin{tabular}{l|cc|rr|rr|rr}
\multicolumn{1}{c|}{XY} & \multicolumn{2}{c}{$\alpha_\mathrm{eq}$
(\AA )}& \multicolumn{2}{c}{$m^\mathrm{X}$ ($\mu_\mathrm{B}$)}&
 \multicolumn{2}{c}{$m^\mathrm{Y}$ ($\mu_\mathrm{B}$)} &
 \multicolumn{2}{c}{$m^\mathrm{total}$ ($\mu_\mathrm{B}$)} \\
  compounds  & LSDA & GGA & LSDA & GGA & LSDA & GGA & LSDA & GGA\\
\hline
MgN & 4.74 & 4.84 & -0.054 & -0.080 & 1.054 & 1.080 & 1.0 & 1.0 \\
MgP & 5.78 & 5.89& \multicolumn{2}{c|}{Non-magnetic}&
\multicolumn{2}{c|}{Non-magnetic}
& \multicolumn{2}{c}{Non-magnetic} \\
MgAs & 5.95 & 6.09& \multicolumn{2}{c|}{Non-magnetic}&
\multicolumn{2}{c|}{Non-magnetic}
 &  \multicolumn{2}{c}{Non-magnetic} \\
MgSb & 6.40 & 6.56& \multicolumn{2}{c|}{Non-magnetic}&
\multicolumn{2}{c|}{Non-magnetic} &
\multicolumn{2}{c}{Non-magnetic} \\ \hline
CaN & 5.31 & 5.45 & -0.053 & -0.073 & 1.053 & 1.073 & 1.0 & 1.0 \\
CaP & 6.39 & 6.56 & 0.059 & 0.025 & 0.941 & 0.975 & 1.0 & 1.0 \\
CaAs & 6.55 & 6.75 & 0.082 & 0.044 & 0.918 & 0.956 & 1.0 & 1.0 \\
CaSb & 7.03 & 7.24 & 0.144 & 0.102 & 0.856 & 0.898 & 1.0 & 1.0 \\
\hline
SrN & 5.66 & 5.82 & -0.071 & -0.093 & 1.071 & 1.093 &1.0 & 1.0\\
SrP & 6.74 & 6.91 & 0.003 & -0.031 & 0.997 & 1.031 & 1.0 & 1.0 \\
SrAs & 6.90 & 7.10 & 0.018 & -0.026 & 0.982 & 1.026 & 1.0 & 1.0 \\
SrSb & 7.37 & 7.60 & 0.065 & 0.018 & 0.935 & 0.982 & 1.0 & 1.0

\end{tabular}
\end{ruledtabular}
\end{table*}

\section{Motivation and computation method}\label{sec3}

From the discussion in Section \ref{sec2} we can conclude that
alkaline-earth metal pnictides are very promising for spintronic
applications since (i) have a small spin magnetic moment per
formula unit (1 $\mu_B$) and thus create small external magnetic
fields, (ii) they present very stable half-metallicity upon
hydrostatic pressure, (iii) their equilibrium lattice constant are
close to a lot of semiconductors, (iv) results on the ZB and RS
structures suggest high values of the Curie temperature, (v) the
half-metallic gaps are wide, and (vi) interfaces with
semiconductors retain half-metallicity. In this article we will
study in detail (Section \ref{sec4}) the electronic and magnetic
properties of several pinctides in the zinc-blende structure
having the chemical formula XY, where the IInd column element X is
Mg, Ca or Sr and the Vth column element Y is N, P, As or Sb. This
study is followed (Section \ref{sec5}) by an extensive
investigation of the properties of these alloys under hydrostatic
pressure, tetragonalization keeping the in-plane lattice
parameters constant and tetragonalization keeping the volume of
the unit cell constant. These latter calculations simulate the
effect of growth on various substrates where one of the three
cases of deformation can occur. To make our study more complete we
have also conducted calculations for RS CaN and SrN under such
deformations.

To perform our calculations we employ the full--potential
nonorthogonal local--orbital minimum--basis band structure scheme
(FPLO)\cite{koepernik} using a dense 20$\times$20$\times$20
\textbf{k}-mesh in the Brillouin zone in the reciprocal space to
perform integrations. Since we are interested in deformations, we
have first to determine the equilibrium lattice constant using
total energy calculations for all compounds under study in the ZB
lattice structure. In Fig. \ref{fig1} we summarize our results
using both the local-spin density (LSDA)\cite{LSDA} and
generalized-gradient (GGA)\cite{GGA} approximations to the
exchange correlation-potential. GGA is know to soften the bonds
with respect to LSDA and to produce equilibrium lattice constants
more close to experimental data (in fact GGA overestimates lattice
constants by less than 1\% while LSDA underestimates them by more
than 1\%). Both LSDA and GGA curves exhibit exactly the same
behavior in Fig. \ref{fig1} with the LSDA lying under the GGA
values as expected. As we move from Mg to Ca and then to Sr for
the same Y element the calculated equilibrium lattice constant
becomes larger since we move from a lighter to a heavier atom. The
same trend occurs for one X-element as we move from N to P and
then to As and Sb. Interestingly the phosphides and arsenides have
very close lying equilibrium lattice parameters. As stated above
the GGA is expected to give a better estimation of the elastic
properties with respect to LSDA while both yield similar
description of the magnetic properties for the same lattice
constant. In Table \ref{table1} we have gathered our results for
all alloys under study using both LSDA and GGA. Our calculated GGA
equilibrium lattice constants for CaP, CaAs and CaSb are in
excellent agreement with the GGA results by Geshi and
collaborators using the full-potential linearized augmented plane
waves methods (FLAPW).\cite{Geshi04} Notice that for MgP, MgAs and
MgSb both approximations converged to a non-magnetic solution. For
all other alloys both LSDA and GGA give a total spin magnetic
moment of 1.0 $\mu_B$ at the LSDA and GGA equilibrium lattice
parameters and thus these alloys are expected to present
half-metallic behavior. The atom-resolved spin magnetic moments
show a small variation in the second decimal digit between GGA and
LSDA. Its origin is the different equilibrium lattice constant
produced by LSDA and GGA, and for the same lattice constant both
LSDA and GGA produce almost identical spin magnetic moments. Thus
in the rest of the paper all presented results have been obtained
using the GGA approximation at the GGA equilibrium lattice
constants. In general our results agree well with the previous
ab-initio studies on these materials discussed in Section
\ref{sec2}.

\begin{figure}
\includegraphics[width=\columnwidth]{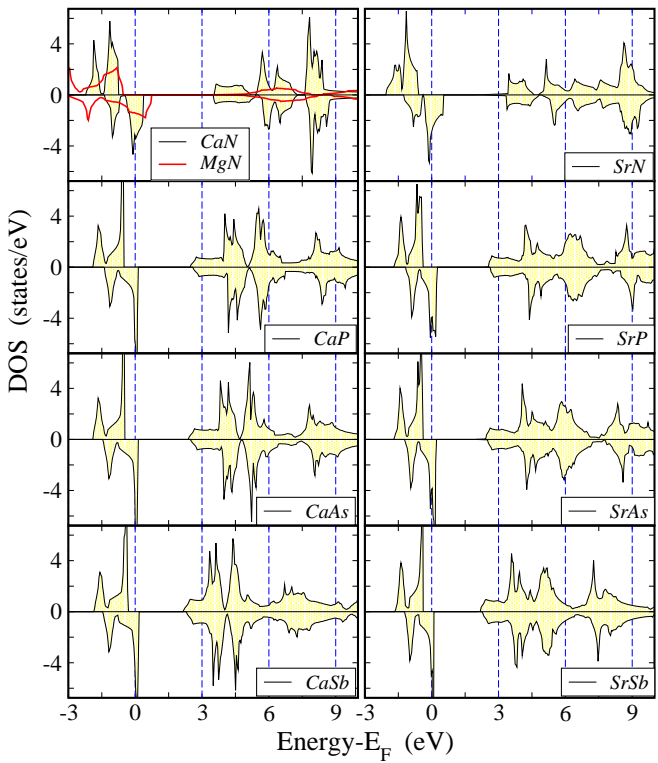}
\caption{(Color online) Total DOS in the primitive cell for all
studied compounds for a wide energy spectrum within the GGA
approximation at the equilibrium lattice constant. The zero energy
has been chosen to represent the Fermi energy (E$_\mathrm{F}$).
The positive DOS values correspond to the majority-spin (spin-up)
electrons and the negative DOS values to the minority-spin
(spin-down) electrons. The antibonding states lying above both the
majority-spin and minority-spin gaps are mainly of cationic
d-character while the states below the gaps are mainly of
anionic-p character as shown in Fig. \ref{fig5}.  Note that we do
present the anionic s-states located at about -10$\sim$-12 eV for
both spin directions.} \label{fig2}
\end{figure}

\begin{figure}
\includegraphics[width=\columnwidth]{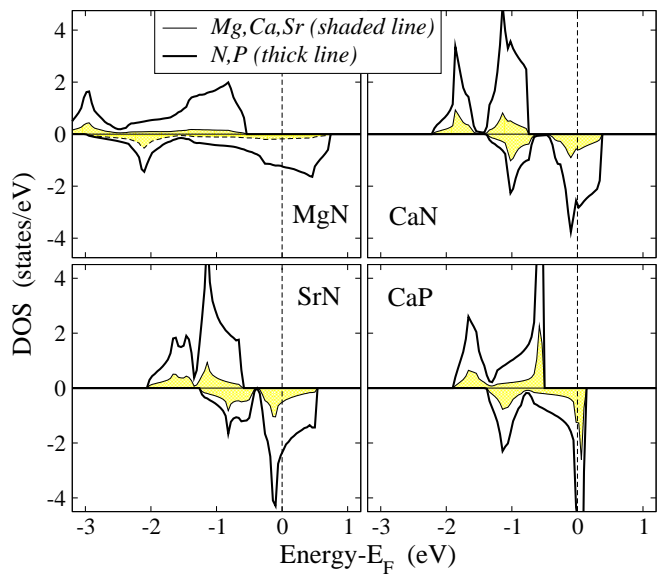}
\caption{(Color online) Atom- and spin-resolved density of states
(DOS) of the four alloys presented also in Fig. \ref{fig4}. We
focus around the Fermi level and do not present the deep-lying
s-states (see caption of Fig. \ref{fig2}) and the high-energy
antibonding states (see Fig. \ref{fig2} for a broader energy
spectrum). Details as in Fig. \ref{fig2}. } \label{fig3}
\end{figure}

 \section{Equilibrium properties}\label{sec4}

As shown in Table \ref{table1} MgP, MgAs and MgSb converge to
non-magnetic solutions while all other alloys seem to be
half-metallic ferromagnets as derived from the Slater-Pauling rule
discussed in Section \ref{sec2} which will be again referred to
later in the text. In Fig. \ref{fig2} we have drawn the total
density of states (DOS) for all magnetic alloys as a function of
the energy setting the Fermi level as the zero energy. Moreover
positive DOS values refer to the majority-spin or spin-up
electrons and negative DOS values to the minority-spin or
spin-down electrons. In all cases we do not represent the
s-valence states which are located at about 10$\sim$12 eV below
the Fermi level. Note that for all compounds under study there are
very large gaps in both majority- and minority-spin electronics
bands structures. The Fermi level crosses the minority-spin states
while it falls within the gap for the majority-spin states and
thus we conclude that these alloys  are half-metallic ferromagnets
where the majority-spin band shows semiconducting behavior while
the minority-spin band shows metallic behavior. As we move from
MgN to CaN and then to SrN the bands become more narrow while as
we move along the Vth column
(N$\rightarrow$P$\rightarrow$As$\rightarrow$Sb) the gap becomes
smaller.

\begin{figure*}
\includegraphics[scale=0.35,angle=270]{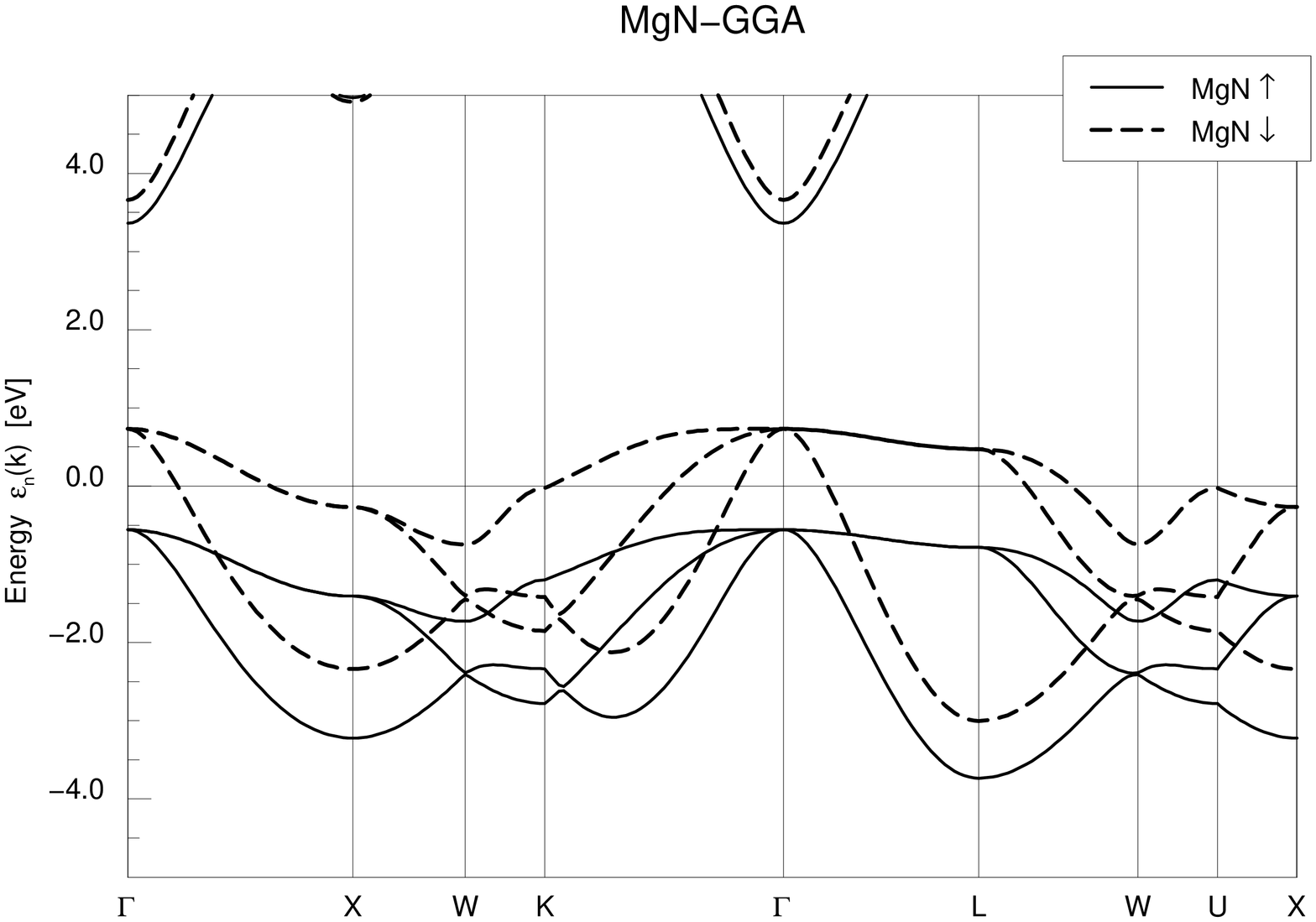}
\includegraphics[scale=0.35,angle=270]{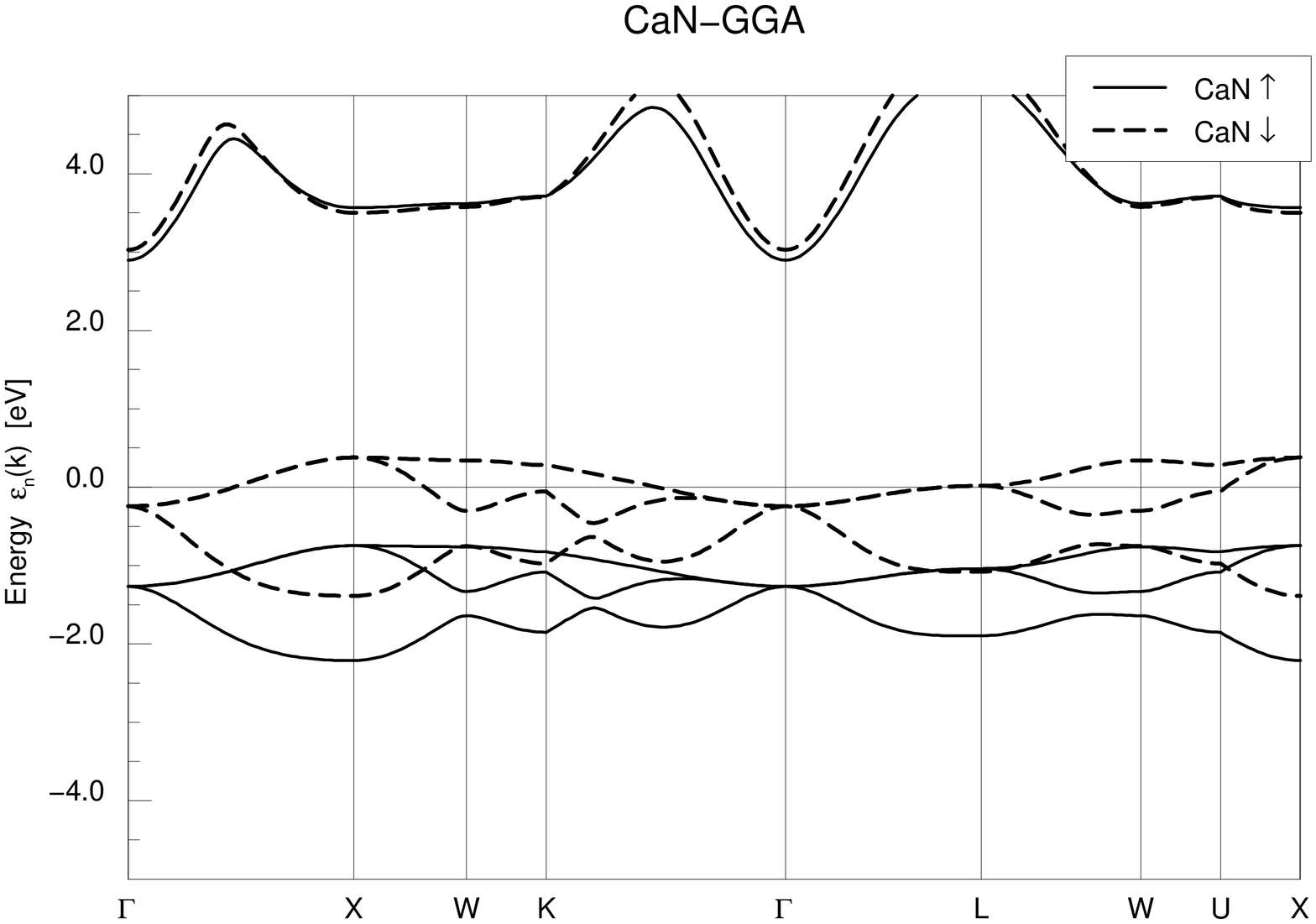}
\vskip 2cm
\includegraphics[scale=0.35,angle=270]{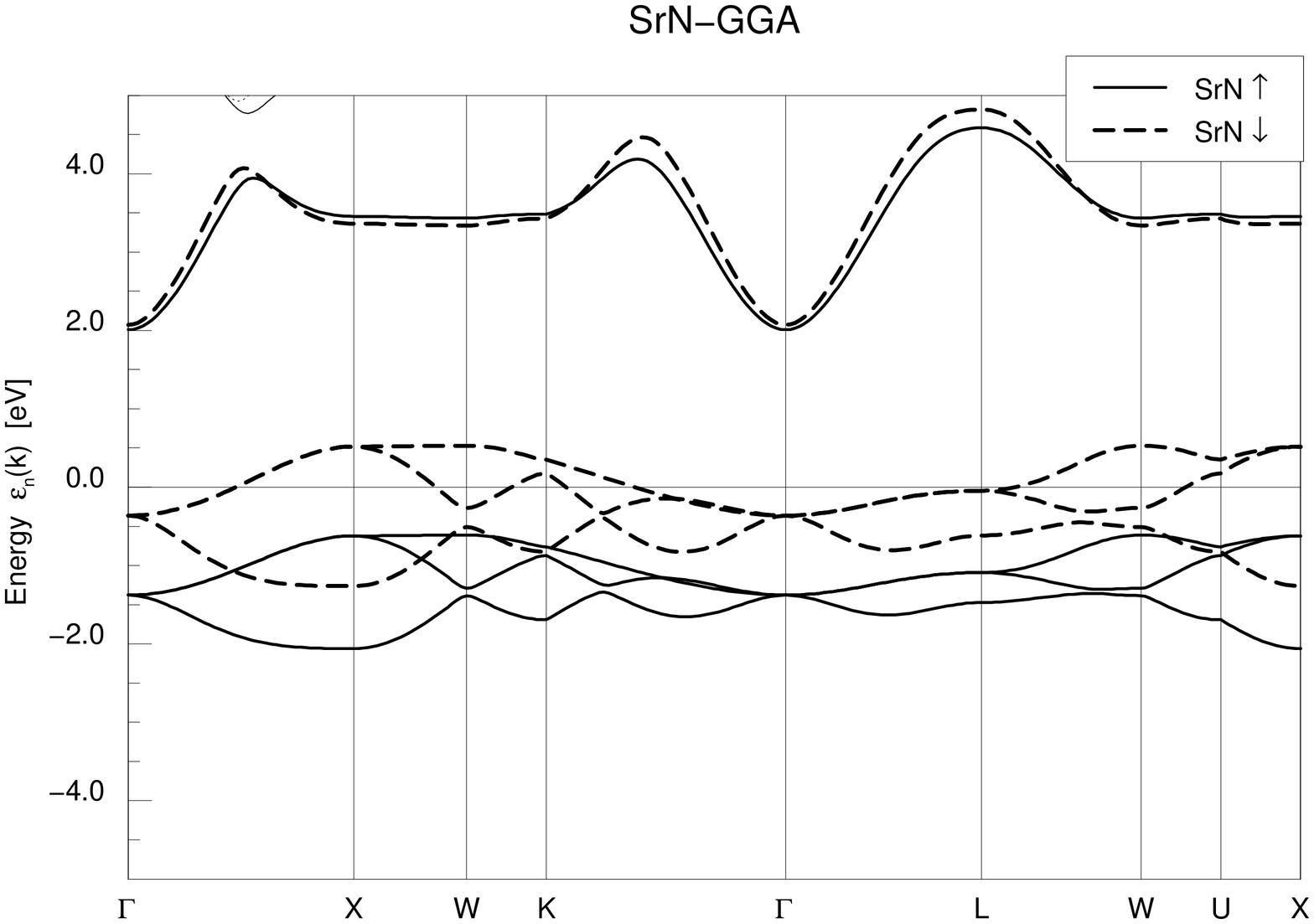}
\includegraphics[scale=0.35,angle=270]{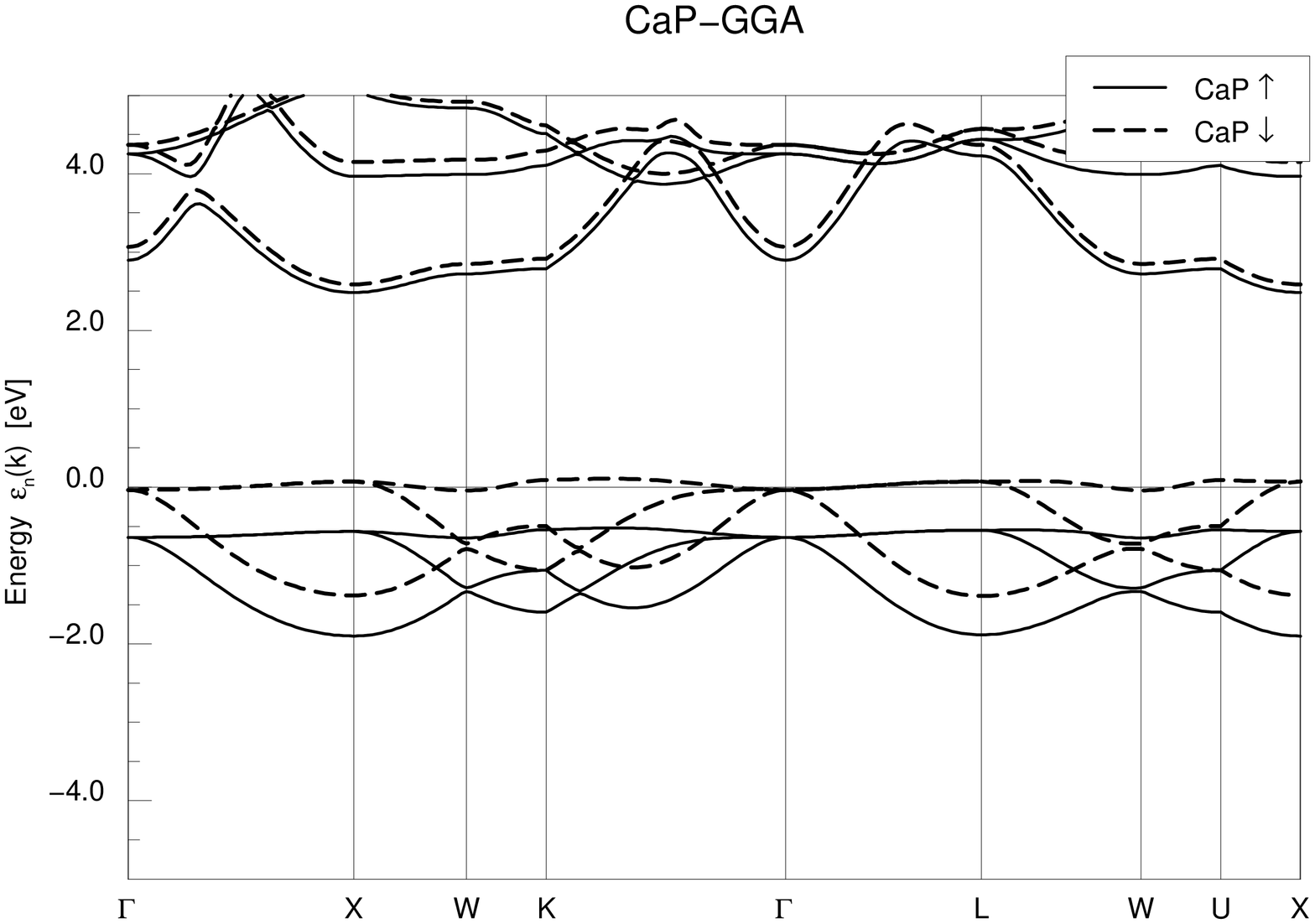}
\vskip 0.5cm \caption{Spin-resolved band structure of MgN, CaN,
SrN and CaP compounds within the GGA approximation along the
high-symmetry lines in the Brillouin zone. The zero energy has
been chosen to represent the Fermi energy. The solid lines
correspond to the majority-spin (spin-up) electrons and the dashed
lines to the minority-spin (spin-down) electrons. The deep-lying
s-band is not shown; it is located at 10$\sim$12 eV below the
Fermi level.}\label{fig4}
\end{figure*}

The main weight of the states lie above the Fermi level. This is
easy to understand if we examine the origin of the gap discussed
in Refs. \onlinecite{Geshi04} and \onlinecite{Kusakabe04}. The
X-elements from the IInd column (Mg,Ca,Sr) provide two valence
electrons (occupying e.g. in Ca the 4s states in the free atom)
while the anions (N,P,As,Sb) provide 5 valence electrons
(\textit{e.g.} in free atom of As the atomic configuration is
4s$^2$ 4p$^3$). In total there are 7 valence electrons per unit
cell. The first two occupy the s-valence states created by the
anions  which lie deep in energy as mentioned above. The p-states
of anions (N,P,As,Sb) hybridize strongly with the unoccupied
triple-degenerated t$_{2g}$ d-states of the cation, which
transform following the same symmetry operations in case of
lattice with tetrahedral symmetry like the zincblende one, and
form bonding and antibonding hybrids which are separated by large
energy gaps. The bonding hybrids contain mostly p-admixture while
the antibonding hybrids are mainly of d-character. The remaining 5
valence electrons occupy the bonding hybrids which are mainly of
anionic p-character in such a way that all three majority-spin
p-states are occupied while in the minority-spin band the Fermi
level crosses the bands so that only the two out of three p-states
are occupied. This gives in total a spin magnetic moment per
formula unit of exactly 1 $\mu_B$. In the case of Ca and Sr the
unoccupied 3d and 4d states, respectively, are relatively close in
energy to the valence p-states of the anion and thus hybridization
is strong enough so that all resulting compounds are magnetic. The
case of Mg should be discussed separately since for light Mg the
unoccupied 3d states are very high in energy with respect to the
occupied states compared to Ca for which also the first unoccupied
d-shell is the 3d one. As a result in MgY alloys the p-d
hybridization is very weak and the resulting bands are very broad.
In MgP, MgAs and MgSb the DOS at the Fermi level is weak and these
alloys prefer to stay non-magnetic and only MgN gains energy when
it becomes magnetic. Sieberer et al using the FLAPW method within
GGA have predicted that MgP and MgAs are magnetic with very small
magnetic moments (0.08 and 0.02 $\mu_B$ respectively) instead of
non magnetic.\cite{Sieberer06} The difference should be attributed
to the fact that, as shown also on Ref. \onlinecite{Sieberer06},
these compounds are at the verge of being magnetic and a small
shift of the predicted Fermi level or a slightly different shape
of the predicted DOS by different ab-initio methods could
eventually lead to the loss of the magnetism,. The latter
compounds can be also seen as a liming case of the Stoner
criterion for transition-metal alloys which states that magnetism
is the ground state only when the DOS at the Fermi level in the
non-magnetic case is quite high and the system gains energy when
being magnetic.

The DOS above the gaps in Fig. \ref{fig2} comes mainly from the
unoccupied d-states and exhibits vanishing spin-splitting. Also
the weight of these states is much larger than for the states
below the gaps since the latter are mainly the p-states (summing
to three per spin direction) while the former are mainly the
d-states (summing to five per spin direction). Thus relevant to
our discussion on the magnetic properties are mainly the p-d
bonding hybrids close to the Fermi level since also the s-states
lying deep in energy are fully occupied (one electron per spin
direction). In Fig. \ref{fig3} we present the atom-resolved DOS
for the three nitrides under study and CaP.  In all cases most of
the DOS is concentrated on the anions reflecting the above
discussion and the cations carry a much smaller weight of the p-d
bonding hybrids. The ratio between the weight at the cation and
the anion site is almost constant among the nitrides but increases
slightly as we move to the CaP (Ca carries a larger weight of the
p-d hybrids in CaP than in CaN). The fact that these states
originate mainly from the p-states of the anions is confirmed
since the curves for the cations DOS follow exactly the shape of
the curves for the anions. This behavior of the DOS is easy to
interpret since the anionic p-states, when expanded around the
cationic centers, become t$_{2g}$-like and thus the cationic d-DOS
in the energy window presented in Fig. \ref{fig3} is a weak
reflection of the anionic p-DOS. Among the nitrides, CaN and SrN
exhibit similar DOS curves but MgN differs considerably since the
hybridization effect for this alloy is much less intense than for
SrN and CaN. Moreover as we substitute P for N in CaN the bands
become more narrow and the exchange-splitting between the
majority- and minority-spin bands decreases (difference in the
energy position of the mass centers of the bands). All these
features can be identified when we look at the band structure for
these four alloys and in Fig. \ref{fig4} we have drawn the
band-structure for both spin-directions in the energy window from
-5 to +5 eV along the high-symmetry axis in the Brillouin zone.
Majority-spin (solid lines) and minority-spin (dashed lines) bands
have exactly the same shape for all four compounds and they are
simply shifted in energy. Similarities are clear if we compare CaN
and SrN where the difference in the energy positions of the bands
is small. In MgN the bands are much broader while in CaP the
exchange splitting is smaller and with respect to CaN more
unoccupied bands appear in the same energy window. As a result of
the reduced exchange splitting also the bonding p-d hybrids are
more narrow. Our calculated band structures for the Ca-based
alloys are in good agreement with the ones calculated by Li and Yu
using a GGA pseudopotential technique.\cite{Li08} Finally we
should note that our band structure confirms that the states below
the gaps are p-d hybrids with mainly p-character, since they are
triple degenerated at the $\Gamma$ point, a characteristic of
either the p or the t$_{2g}$ states in the lattices of tetrahedral
symmetry likes the zincblende one.

\begin{figure}
\includegraphics[width=\columnwidth]{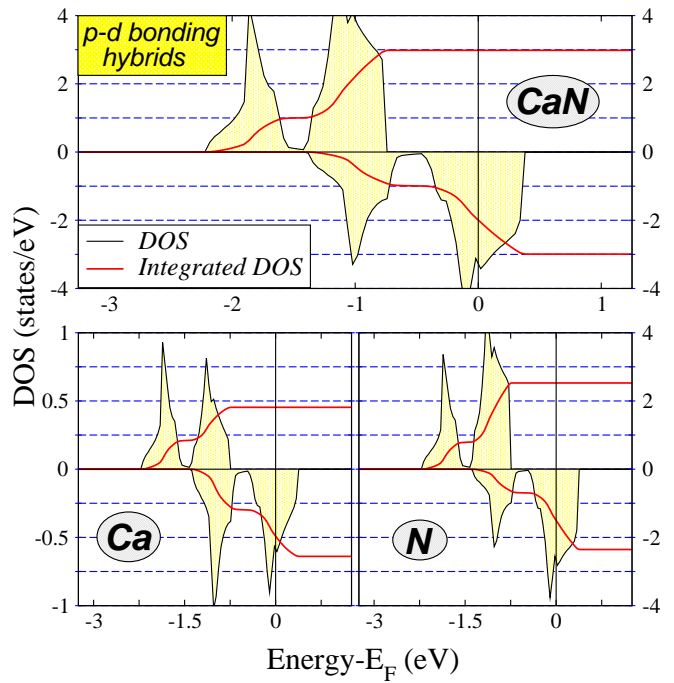}
\caption{(Color online)  Spin-resolved DOS of the p-d bonding
hybrids (shaded line) and the corresponding integrated DOS (thick
red line) for CaN within the GGA approximation and its projection
on the Ca and N atoms. We do not present the s-states which are
deep in energy and contain one electron per spin-direction with
their weight almost exclusively at the N-sites. Notice that the
p-d hybrids can accommodate 3 electrons per spin direction. All
three majority-spin states ($\sim$2.5 at the N site and $\sim$0.5
at the Ca site) are occupied, while only two out of three
($\sim$1.5 at the N site and $\sim$0.5 at the Ca site) are
occupied resulting in a total spin magnetic moment of 1
$\mu_\mathrm{B}$.} \label{fig5}
\end{figure}

In the last paragraph we shall also discuss the spin magnetic
moments presented in Table \ref{table1}. The cations carry very
small spin magnetic moments of less than $|0.1|\:\mu_B$ and almost
all spin moment is carried from the anions. The spin magnetic
moment reflects directly the imbalance in the total number of
occupied states. In Fig. \ref{fig5} we have drawn both the DOS and
the integrated DOS for the bonding p-d hybrids in the case of CaN.
We do not include in the integrated DOS the deep-lying s-states
which accommodate one electron per spin direction and thus carry
no net spin moment. If we integrate the total DOS we have three
states per spin direction as expected by the band-structure
discussed just above. All the majority-spin states are occupied
while the integrated DOS line cross the Fermi level at the value
of two. Thus in the minority-spin band only the two out of three
states are occupied and the total spin magnetic moment is 1
$\mu_B$ since we have an imbalance in the electronic charge
counting of one. This picture agrees with the Slater-Pauling
behavior for the sp-electron ferromagnets which we have discussed
in Section \ref{sec1}. If we now look at the Ca resolved DOS in
total we have slightly less than 0.5 electrons in the
majority-spin band and about 0.65 in the minority-spin band. All
majority states are occupied while in the minority-spin band only
the 0.5 out of 0.65 are occupied. This small imbalance in the
electron counting leads to a small negative spin magnetic moment
at the Ca site of -0.073 $\mu_B$. Small variation of the values of
the  cationic integrated DOS at the Fermi level leads to small
variation of the cationic spin magnetic moment around zero. At the
N site we have about 2.5 states in the majority-spin band which
are completely occupied while in the minority-spin bands we have
about 2.35 states in total but only $\sim$1.5 are occupied leading
to a N spin magnetic moment of about 1 $\mu_B$. Previous
calculations led to a similar picture of the atom resolved
spin-magnetic moments with the cation spin moments being one order
of magnitude smaller than the anion spin magnetic
moments.\cite{Geshi04,Laref11} Thus to conclude, we should have in
all cases a total spin magnetic moment of 1 $\mu_B$ since we have
seven valence electrons, all compounds are half-metals and the
atom resolved spin moments fluctuate around 0 or 1 $\mu_B$ for
cations and anions, respectively, depending on the p-d
hybridization intensity.

\begin{figure*}
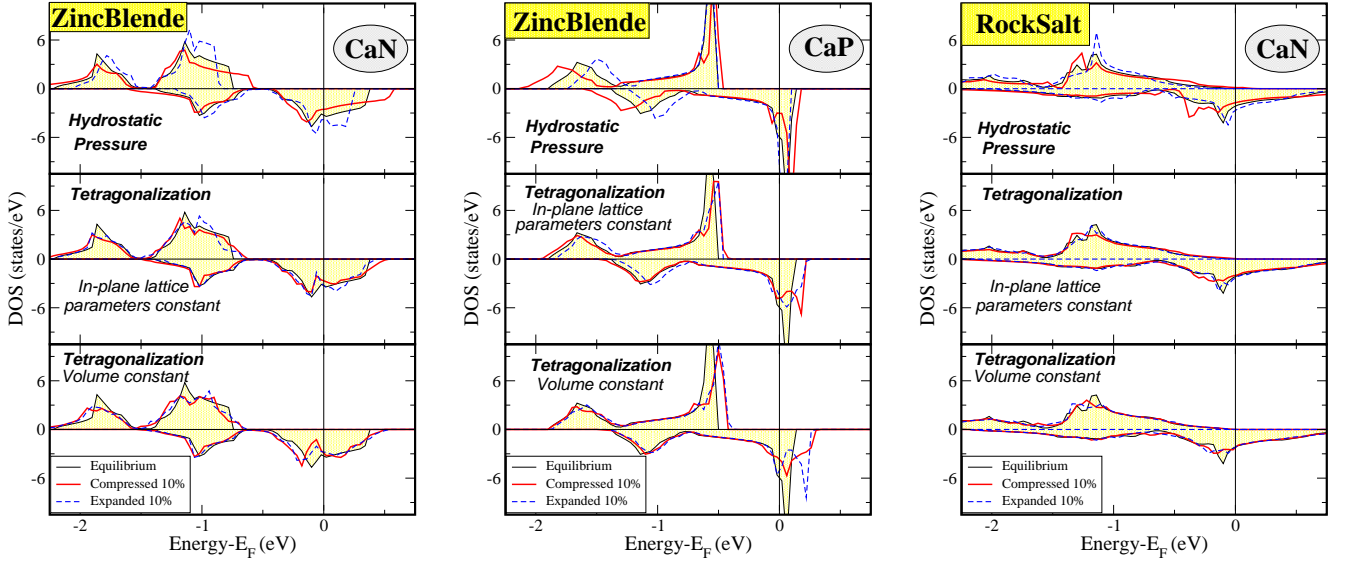

\includegraphics[scale=0.3]{fig6a.eps}
\hskip 0.5cm
\includegraphics[scale=0.3]{fig6b.eps}
\hskip 0.5cm
\includegraphics[scale=0.3]{fig6c.eps}
\caption{(Color online) Total DOS of CaN (left panels) and CaP
(middle panels) compounds in the zincblende structure and CaN
(right panels) in the rocksalt structure within GGA at the
equilibrium lattice constants compared to : (upper panel) 10\%
uniform expansion or contraction of the lattice as in the case of
hydrostatic pressure, (middle panel) tetragonalization keeping the
in-plane lattice parameters equal to the equilibrium ones and
expanding or contracting the out-of-plane lattice parameter by
10\%, and (lower panel) tetragonalization expanding or contracting
the in-plane lattice parameters by 10\% and in the same time
changing accordingly the out-of-plane lattice parameter to keep
the unit cell volume constant to the equilibrium one. }
\label{fig6}
\end{figure*}

\begin{table*}
\caption{Calculated spin magnetic moments (in $\mu_B$) for CaN and
CaP in the zincblende structure and CaN in the rocksalt structure
using the FPLO method within the GGA approximation. The first line
corresponds to the ideal equilibrium lattice constants and we have
considered three cases of deformation: (i) hydrostatic pressure
(Hydro) which is a uniform compression and expansion of the cubic
lattice and the percentage in the second column shows the change
in the lattice constant ("-" means compression and "+" means
expansion), (ii) tetragonalization of the lattice keeping the
in-plane lattice parameters constant and equal to the equilibrium
values and changing only the lattice parameter along the c-axis
(Tetra-I) and the percentage in the second column shows this
change, and (iii) tetragonalization where we vary the in-plane
lattice parameters by the percentage shown in the second column
and in the same time we vary also the lattice parameter along the
c-axis so that the unit-cell volume is kept equal to the
equilibrium one (Tetra-II). In all cases the total spin-magnetic
is kept equal to 1.0 $\mu_B$ (even in the case of $\pm$10\%\
hydrostatic pressure where the unit-cell volume changes by almost
30\% ) and thus the half-metallicity is preserved. }
\label{table2}
\begin{ruledtabular}
 \begin{tabular}{lr|rrr|rrr|rrr}
&& \multicolumn{3}{c}{CaN-ZincBlende (5.45 \AA )}
 &\multicolumn{3}{c}{CaP-ZincBlende (6.56 \AA )} &             \multicolumn{3}{c}{CaN-RockSalt (5.02\AA )}\\
     & Case & Ca & N & Total & Ca & P & Total &               Ca & N & Total\\\hline
 Ideal &  & -0.073 & 1.073 & 1.0 & 0.025& 0.975& 1.0         & -0.065& 1.065& 1.0  \\ \hline
Hydro & -1\% &  -0.072&1.072 & 1.0& 0.025 & 0.975 & 1.0          & -0.061 &1.061 & 1.0\\
& -5\%  & -0.071& 1.071&1.0& 0.025 & 0.975& 1.0            & -0.059 & 1.057& 0.999\\
& -10\%  &-0.068 &1.068 &1.0& 0.026 & 0.974& 1.0                  & -0.048 & 1.017 & 0.969\\
&+1\%   & -0.074 & 1.074&1.0& 0.025 & 0.975 &1.0                & -0.067 & 1.067 &1.0 \\
&+5\% & -0.075& 1.075&1.0& 0.026 & 0.974& 1.0                   & -0.070& 1.070& 1.0\\
& +10\%  & -0.080 &1.080 & 1.0& 0.026 & 0.974  &1.0    & -0.079&
1.079&1.0 \\ \hline
Tetra-I & -1\%  & -0.072&1.072 &1.0&  0.025 & 0.975& 1.0          & -0.063& 1.063& 1.0\\
& -5\%   & -0.072& 1.072&1.0& 0.025 & 0.975 & 1.0           & -0.062 & 1.062 & 1.0 \\
& -10\%   & -0.071&1.071 &1.0& 0.026 & 0.974& 1.0                & -0.058 & 1.055& 0.997 \\
&+1\%  & -0.073& 1.073&1.0&  0.025 & 0.975& 1.0                & -0.065& 1.065& 1.0\\
&+5\% & -0.073& 1.073&1.0& 0.026 & 0.974& 1.0                    & -0.066& 1.066& 1.0\\
& +10\%  &-0.074 &1.074 &1.0& 0.027 & 0.973& 1.0       & -0.068&
1.068& 1.0\\ \hline
Tetra-II & -1\%  & -0.073 &1.073 &1.0& 0.026 & 0.974&1.0         & -0.064& 1.064&1.0 \\
& -5\%   & -0.072 &1.072 &1.0& 0.027 & 0.973&1.0        & -0.064& 1.064&1.0 \\
& -10\%  & -0.071 &1.071 &1.0& 0.027 & 0.973& 1.0                & -0.063& 1.063& 1.0\\
&+1\% & -0.073 &1.073 &1.0 &0.026 & 0.974& 1.0                 & -0.064& 1.064& 1.0\\
&+5\% & -0.072 &1.072 &1.0& 0.027 & 0.973 &1.0                  & -0.064 &1.064 &1.0 \\
& +10\%   & -0.071 &1.071 &1.0&0.029 & 0.971 &1.0 &-0.062 &1.062
&1.0
\end{tabular}
\end{ruledtabular}
\end{table*}

\section{Deformations}\label{sec5}

To simulate deformation of the lattices  with respect to
equilibrium we took into account three cases: (i) hydrostatic
pressure (Hydro) which is a uniform compression and expansion of
the cubic lattice, (ii) tetragonalization of the lattice keeping
the in-plane lattice parameters constants and equal to the
equilibrium values and changing only the lattice parameter along
the c-axis (Tetra-I), and (iii) tetragonalization where we vary
 both the in-plane and out-of-plane parameters keeping the unit
cell volume constant to the equilibrium (Tetra-II). In Table
\ref{table2} we have gathered our results concerning the spin
magnetic moments for CaN and CaP adopting the ZB-lattice. We have
also performed similar calculations for SrN and SrP but the
behavior of the Sr compounds is identical to the Ca ones and thus
we do not present them. Moreover we present results also for CaN
in the rocksalt structure which has been also widely studied in
literature. Preliminary results on SrN in both ZB and RS structure
taking into account only the case of tetragonalization keeping the
unit cell volume constant have been published in Ref.
\onlinecite{JAP}. The latter type of distortion is expected to
occur more often when these compounds are grown on top of various
substrates, but the other two cases under study are also
susceptible to occur and we include them aiming to give a better
description of the properties of these alloys under various
distortions. In the second column of the table we present for each
one of the three cases the percentage of change; "-" means
compression by that percentage and "+" means expansion. Moreover
in the case of the hydrostatic pressure the percentage refers to
the lattice constant since the expansion or compression is
uniform, in the case of Tetra-I where we keep constant the
in-plane lattice parameters the percentage refers to the
out-of-plane lattice parameter, and finally in case Tetra-II where
we keep the unit-cell volume constant the percentage refers to the
in-plane lattice constants. For each case we took six values of
the percentage $\pm$1 \%, $\pm$5 \% and $\pm$10 \%. The larger
changes in the unit cell volume occur for the Hydrostatic case
where $\pm$10 \%\ change of the lattice constant means compression
or expansion of the unit cell volume by about 27 \%. Amazingly
even for these extreme cases CaN and CaP in the ZB lattice remain
half-metallic with a total spin magnetic moment of 1 $\mu_B$. CaN
in the RS-lattice shows similar behavior to the ZB one and only in
the case of 10\% compression it is at the verge of
half-metallicity and Fermi level is located at the edge of the
occupied majority-spin states; this behavior occurs because the
half-metallic gap (energy difference between the upper edge of the
occupied majority-spin states and the Fermi level) is 0.1 eV much
smaller than the compounds in the ZB lattice. Deformations in CaN
lead to small changes of the absolute values of the Ca and N
atomic spin magnetic moments of less than 0.07 $\mu_B$ in such a
way that they cancel each other keeping the total spin moment
constant. When we replace P for N in CaN the fluctuations of the
atomic spin magnetic moments of the constituent atoms are even
smaller.

In Fig. \ref{fig6} we present for the alloys in Table \ref{table2}
the total DOS taking into account expansion or compression by 10\%
which are the limiting cases studied and compare them with the DOS
for the ideal cubic lattices at the equilibrium lattice constants.
We focus on the energy window including the p-d bonding hybrids
which are the most relevant for magnetism. In the case of the
$\pm$1\% and $\pm$5\% cases the DOS for the distorted lattices
coincide with the DOS for the equilibrium lattice and the curves
fall almost one on top of the other; the only noticeable
difference concerns the hydrostatic pressure case where
compression leads to slightly wider bands and thus smaller pick
intensities. Differences in the DOS are only visible in the case
of the $\pm$10\% deformations presented in the figure. The most
important changes occur in the case of hydrostatic pressure where
compression leads to wider bands and expansion to narrower bands,
this behavior being typical for p-orbitals. In the case of
tetragonalization the ZB-CaN shows very small changes in the DOS
shape while for ZB-CaP we remark a small redistribution of the
weight of the empty minority-spin states just above the Fermi
level. For CaN in the RS-lattice the changes in the DOS are even
more marginal than for the ZB-lattice. The large half-metallic
gaps especially in the case of the ZB lattice ensure that any
small change in the characteristics of the DOS does not affect the
half-metallicity and the Fermi level still crosses exclusively the
minority-spin bands. The same image is also valid for the SrN and
SrP alloys. Thus we can safely conclude that deformation of the
lattice only marginally affects the electronic properties as was
the case also for the magnetic properties discussed just above.

\section{Conclusions}\label{sec7}

We have studied the properties of the XY (X= Mg, Ca, Sr and Y= N,
P, As, Sb) pnictides crystallizing in the zincblende structure
using a full-potential electronic structure method in conjunction
to the generalized gradient approximation. We have shown that the
Ca- and Sr-based alloys should be classified as sp-electron
ferromagnets (also known as $d^0$-ferromagnets). Moreover they are
half-metals since the Fermi level crosses only minority-spin
states and the total spin magnetic moment per formula unit was
found to be 1.0 $\mu_B$. Contrary to the Ca and Sr alloys, in the
case of the Mg alloys the p-d hybridization effect is much weaker
and only MgN is a half-metallic ferromagnet while MgP, MgAs and
MgSb prefer energetically to be non-magnetic. We have also studied
for these alloys the effect of deformation taking into account
both the cases of hydrostatic pressure and tetragonalization
keeping constant either the in-plane lattice parameters or the
unit cell volume. Even large degrees of deformation only
marginally affected the electronic and magnetic properties of
these alloys which kept their half-metallic ferromagnetic
character. Similar behavior is found also when they crystallize in
the rocksalt instead of the zincblende structure. Our results
suggest that the alkaline-earth metal monopnictides are promising
materials for spintronic applications.

\end{document}